\documentclass[runningheads]{llncs}
\usepackage{graphicx}
\usepackage{amsmath}

\usepackage{xcolor}
\usepackage{tikz-qtree}

\begin{document}
\title{Query Generation for Patent Retrieval with Keyword Extraction based on Syntactic Features}
\titlerunning{Patent Claims Keyword Extraction}
%
%
\author{Julien Rossi\inst{1} \and \\
	Matthias Wirth\inst{2} \and \\
Evangelos Kanoulas\inst{3}} 
\authorrunning{J. Rossi \& al}
%
\institute{Amsterdam Business School, University of Amsterdam \and European Patent Office \and Institute of Informatics, University of Amsterdam}
\maketitle              
\begin{abstract}
This paper describes a new method to extract relevant keywords from patent claims, as part of the task of retrieving other patents with similar claims (search for prior art). The method combines a qualitative analysis of the writing style of the claims with NLP methods to parse text, in order to represent a legal text as a specialization arborescence of terms. In this setting, the set of extracted keywords are yielding better search results than keywords extracted with traditional methods such as tf-idf. The performance is measured on the search results of a query consisting of the extracted keywords.
\end{abstract}
\section{Introduction}
A patent is an economic monopoly for using or selling a patented invention, it is the mission of the Patent Offices to properly identify innovation and not grant undue protection. The search for prior art is a critical task in the processing of a patent filing. A patent examiner has to establish whether or not the claims of novelty made in the filing stands in view of the prior art.

A patent filing consists of four different sections: Metadata, Abstract, Description and Claims. If needed, drawings may be added. The Description is a technical text describing the invention in technical details, with the potential support of drawings or tables. The Claims provide a legal description of the scope of the invention for which the applicant seeks patent protection. Claims are hierarchically structured, with \textit{dependent claims} referring and detailing other claims, while \textit{independent claims} do not refer other claims. For more details regarding the patent filing's inner structure, and the task faced by professionals, we refer the reader to Lupu and Hanbury \cite{Lupu:2013:PR:2688159.2688160} and Joho et al. \cite{enlighten40041}.

This work focuses on improving the effectiveness of search for prior art (or invalidity search). This is the search performed by patent examiners to establish the rightfulness of the claims in a patent filing (or application). It is a high recall task, where the target is to retrieve from databases a list of documents that relate to the invention and demonstrate whether the invention at hand is indeed novel or not. These documents serve as a context for the invention, and support the opinion of patent examiners with regards to the validity of the claims of an application. The list of retrieved and annotated documents is compiled in the Search Report, and it is further communicated to the patent applicant.

A number of methods have been proposed in the literature with the purpose of searching for prior art without the need to construct a Boolean Query, based on automated keyword-based query generation. Research in this area has been supported by two evaluation forums, NTCIR \cite{Lupu2017} and CLEF-IP \cite{CLEFIP}. Many of the submitted systems were based, partially or entirely, on keyword extraction out of the description or the claims. Konishi \cite{Konishi05} extracted topics of the invention together with description of these topics by matching lexical patterns, while Golestan et al. \cite{GolestanFar:2015:TST:2766462.2767801} used BM25 to extract the patent query, and later used the first relevant document pointed by the user to reduce the query to its most relevant terms. Mase et al. \cite{Mase:2005:PTP:1105696.1105702} uses TF and TF-IDF to identify important words in claims and weigh them, although it makes the remark that claims are highly redundant and therefore a poor fit for term frequency techniques. Similar techniques based on word frequency, using either BM25 or TF-IDF, were also used in Lopez and Romary \cite{lopez:hal-00411835,lopez:inria-00510267} and Verberne and d’Hondt \cite{2010CLEFVerberne}.

In this work we also follow the methodology of keyword-based query generation, but different from previous work we do not rely on term frequency to identify semantic saliency. Instead we depend on creating a claim tree of dependent claims out of the claim section of the patent and a specialization tree of terms used in claims, and using these trees and their structure to identify novelty terms.

Suzuki and Takatsuka~\cite{C16-1113} tackled the problem of identifying novelty-related keywords based on claims in an Information Extraction task. Their method relies as well on the consideration that claims are written in descriptive way, where salient details appear later in the text. Their system is mostly rule-based and identifies typical formulation of specialization in Japanese. The capacity of those extracted keywords to represent novelty in the claim text is evaluated based on the differences in claims keywords between the filing and the granting of the patent. The patent process allows for amendments in claims during the process, under the constraint of not extending the protection domain.

In this work, we propose a more generic method based on a constituency parser and chunking, that does not rely on specific keywords or sentence construction, but rather on its underlying grammar to extract signals of semantic importance of words. This method does not rely on term-frequency as an indicator of semantic saliency. We show that this method can generate queries yielding to better search results, and we evaluate improved recall but also improved ranking.

To summarize, in this work we attempt to answer the following research question: \textit{Can we improve the retrieval of relevant document by selecting keywords based on syntactical signals of semantic importance, rather than term-frequency?}

\section{Methodology}
In a nutshell our method works as follows: (1) Based on the complete claim set of a patent filing we generate a claim tree; (2) we then generate a specialization tree for each claim, and (3) score words based on their appearances in tree nodes; (4) we then submit the selected top-n keywords to a search engine. The search engine is the ad-hoc Lucene production instance of the EPO, with undisclosed modified similarity, based on TF-IDF. 

\subsection{Generating the Claim Tree}
The dependencies between claims are written in free-form language. For example, ``The lubricant concentrate according to claim 3'', ``The lubricant concentrate according to any one of claims 3 to 5'', ``The lubricant concentrate according to any one of claims 3 – 6''. Individual claims are already provided as separate numbered texts. We used a small set of undisclosed regular expression to identify the dependencies and create the claim trees. \footnote{Strictly speaking, these are not trees, since each claim has an identified list of parent nodes.}

Using the Punkt Tokenizer, we identify that only 0.22\% of all claims are split into multiple sentences. Therefore, for simplicity, we considered each claim as a single sentence. Each claim is parsed by the Stanford Core NLP Constituency Parser \cite{P14-5010}, which provides the word tokenization, the POS tagging and the constituency parsing itself. Because of the unusual language of the claims, it is typical that words like ``said'' or ``claim'' are misclassified as verbs where they are instead used as relative adjectives. For example, in a sentence such as ``(...) characterized in that said sectors are buried (...)'', ``said'' is tagged as Verb, which leads the constituency parser to detect Subject and Complement. The incorrect POS tagging of ``said'' has repercussions, such as the creation of a VP chunk. The correct POS tagging for ``said'' is ``JJ'' in the Penn Treebank Tag set. We created a new annotator in the Stanford Core NLP Server, which is bypassing the POS tagger, and feeding the POS tags directly to the constituency parser, so that we can re-parse the sentence.
Now the parsing is changed and reflects better the grammatical and semantic structure of the whole sentence. This correction was proposed by Hu et al. \cite{DBLP:journals/corr/HuCW16}, this paper re-used their dataset to build a classifier SVM, using word2vec \cite{Mikolov2013} embeddings as the word features. For each word with the POS tag of a verb, the 3-gram with the verb as the middle word is used as an input to the SVM. The classification indicates if in this context, the word should be tagged as a verb or as something else. We reparse each sentence after retagging each verb.

\subsection{Generating Specialization Trees}
The constituency parsing tree is traversed depth-first, creating a string of tags, both POS and Chunks tags. The set of regular expression we use identify two types of patterns:
\begin{itemize}
	\item Composition : “a system comprising this and that”, the constituency parsing allows for a lot of flexibility in the actual wording, as the chunk structure stays stable in that situation.
	\item Specialization : “a system made of this, which is on top of this and that”, again the chunk structure is very stable and resistant to the diversity of the wording. 
\end{itemize}

We observe that the chunking produced by the Stanford Parser is very stable over the actual phrasing and choice of verbs, words and delimiters. It efficiently reduces lexical and morphological variations of the concepts of composition and specialization to a few chunk patterns. The text is also now reduced by removing stopwords, as they are relevant for parsing but not carrying information. We leverage the chunking stability to then fold sentences into trees. Each node of that tree will contain a part of the sentence and the tree is built as follows:
\begin{itemize}
	\item Composition : Head node contains “a system”, attached to two child nodes “this” and “that”
	\item Specialization : Head node contains “a system”, attached to one child node “this”, itself attached to 1 child node “this and that”
\end{itemize}

As an example here is Claim 37 from patent filing EP-1748300-A1:
\textit{“Method according to one or more of the preceding claims 25 to 36, characterized in that initial iteration steps for determining compensation dipoles by means of quadratic or linear programming can provide in combination a modification for each subsequent iteration step consisting in a reduction of constraints such that the partial solution converges progressively towards a solution that is considered an optimum one.”}

This text produces the following tree:\\

\noindent\makebox[\textwidth]
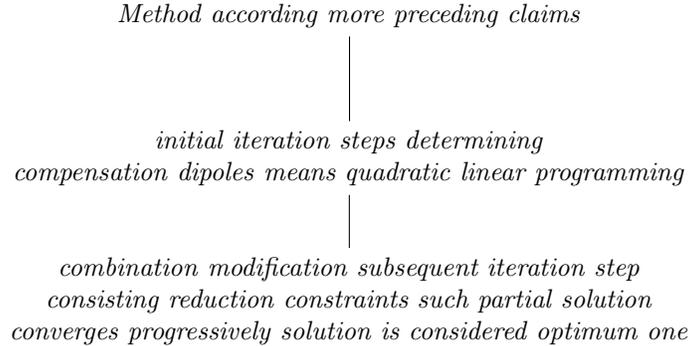
{
\begin{tikzpicture}
\tikzset{align=center,level distance=60}
\Tree [.{\textit{Method according more preceding claims}} 
			[.{\textit{initial iteration steps determining} \\ \textit{compensation dipoles means quadratic linear programming}} 
				[.{\textit{combination modification subsequent iteration step} \\ \textit{consisting reduction constraints such partial solution} \\ \textit{converges progressively solution is considered optimum one}} ] ] ]
\end{tikzpicture}
}
\newline
\newline 
The specialization tree is the representation of one claim as a tree based on the relations of composition and specialization  between chunks of the text. We identify words \(w\) as belonging to specialization tree nodes \(n_{i}\):
\begin{equation}
  \forall w,\:N(w)=\{n_{i},\:where\:w \in n_{i}\}
\end{equation}
\begin{equation}
  \forall w,\:P(w)=\{ \left(nd\left(n_{i}\right),\:nh\left(n_{i}\right),\:cd\left(n_{i}\right)\right),\:n_{i}\in N\left(w\right) \}
\end{equation}

$nd$ is the depth of the node within the specialization tree, $nh$ is the height of the node within the specialization tree, $cd$ is the depth of this claim within the claim tree of that patent. We use depth and height in their standard definition from the context of trees.

\subsection{Scoring keywords}
Words are then grouped by stem, using the Porter Stemmer. For a specific stem, we record which one of all the words with the same stem had the highest number of occurrences.
\begin{equation}
\forall stem\:s,\:P\left(s\right)=\bigcup_{stem\left(w\right)=s}P\left(w\right)
\end{equation}

The scoring method has to favor words that are located deep within the specialization tree, as they relate to finer details of the invention, which is where we expect an invention to stand out of other similar inventions. We also want the scoring to favor words that are within claims that are deep into the claim tree, for the same reason as above, as a claim discloses finer details about the claims it depends on.

We devised two scoring methods:\\
\begin{align}
CLST05\left(s\right) &=\sum_{P\left(s\right)}e^{\alpha_{05}*\frac{nd}{nd+nh-1}+\beta_{05}*cd}\\
CLST06\left(s\right) &=\sum_{P\left(s\right)}e^{\alpha_{06}*\max\left(nd\right)+\beta_{06}*\max\left(cd\right)}
\end{align}

The hyperparameters \(\alpha_{05}, \alpha_{06}, \beta_{05}, \beta_{06}\) are determined by a grid search.

The top-n stems with the highest scores are selected to construct a query, which is the concatenation of the words associated with these stems. We experimented with the number of words in the query to establish what the best result is, as it is a trade-off between having a few powerful keywords but not enough diversity, or diverse keywords but retrieving too much noise. 

\section{Experimental Setup}

\paragraph{Dataset.} We used the CLEF-IP 2011 Topic Collection as a basic dataset. This collection contains patent documents in all 3 languages of the European Patent Office (EPO), i.e. English, French, and German, and the labels to identify the definitive list of relevant documents for each case. We can then compare our search results with the Gold Standard. The search database is the complete historical worldwide repository of patents. The setting is to search for relevant documents based on the claims from the seed document, searching through the claims of the documents in the corpus.

\paragraph{Search Engine.} We used the Lucene search engine. As a reminder, here is the Query Terms and Documents similarity score used by Lucene, for a query $q$ and a document $d$\footnote{\url{https://lucene.apache.org/core/3\_5\_0/api/core/org/apache/lucene/search/Similarity.html}}:
\begin{equation}
\label{lucene}
score(q,d) = coord(q,d)*queryNorm(q)*\sum_{t \in q}\left(tf(t \in d)*idf(t)^2*boost(t)*norm(t,d)\right)
\end{equation}

\paragraph{Baseline.} The baseline is a system developed at the EPO, known under the MLT acronym. It is configured to act within the same parameters, using the claims as source for query, and the claims within the search database as corpus. It extracts 70 keywords from the source document claims, using TF-IDF, and uses these keywords as a query for the Lucene search engine. It is calibrated for the specific task of retrieving patent documents based on the contents of claims only. We use it as it is representative of related work that uses TF-IDF term weighting to extract keywords.

\paragraph{System Configuration.} Our methods are called “CLST-05” and “CLST-06”, and each method has three variants: ``As is'', ``BOOST'' where the word scores are used as boost factors for the search engine, and ``NO-RETAG'' where the POS tagging is not corrected. The boost increases the scoring of a document that contains the boosted terms. It indicates a hierarchy in the importance of keywords, as shown in Equation \ref{lucene}.  We used the Lucene search engine in place at the EPO, which works against search database. This search engine receives our query and returns a list of ranked search results. Each result is a family of patent documents.

\paragraph{Evaluation Metrics.} We evaluate the performance of the different keyword-based query generation methods on the basis of Recall@100, and PRES@100, similar to the evaluation performed under CLEF-IP.
PRES is a metric introduced by Magdy and Jones \cite{4174c8be25f0425e893387123b9df672}, aiming at a better evaluation of the performance of a retrieval system in a setting with only a few actually relevant documents among many. It is a correction of the Recall, the metric decreasing when relevant documents appear further down the search results.
\begin{align}
PRES &=1-\frac{\frac{\sum{r_{i}}}{n}-\frac{n+1}{2}}{N_{max}}, \text{where} \\
\label{incorrect} \sum r_{i} &= \sum_{i=1}^{nR}r_{i} + nR\left(N_{max}+n\right)-\frac{nR\left(nR-1\right)}{2}
\end{align}
Where \(N_{max}\) is the cut at which we compute PRES, \(n\) is the number of relevant documents in the collection for the seed document, \(R\) is the recall (so \(nR\) is the number of relevant documents retrieved before rank \(N_{max}\)). \(sum{r_i}\) is intended as the sum of the ranks at which relevant documents appear, with all relevant documents retrieved after rank \(N_{max}\) considered as ranked at positions \(N_{max}+1, N_{max}+2, \ldots \).

We had to correct the formula presented in the original paper as it was producing results out of the  range \(\left[0,1\right]\). We used equation \ref{correct} for our calculations.

\begin{align}
\label{correct} \sum r_{i} &= \sum_{i=1}^{nR}r_{i} + \sum_{i=nR+1}^{n}{\left(N_{max}+n-\left(i-nR-1\right)\right)}
\end{align}

\section{Results and Analysis}
In the first place, we can compare the average PRES@100 and Recall@100 to the baseline. We developed 6 different systems and evaluated them. For each system we can also select how many words we extract as keywords, from 10 to 100 by increment of 10. The performance of all those systems is given in Tables~\ref{PRES@100} and \ref{RECALL@100}, for PRES@100 and Recall@100, respectively.
\setlength{\tabcolsep}{5pt}
\begin{table}
  \caption{PRES@100}
  \label{PRES@100}
  \centering
  \begin{tabular}{|l|llllllllll|}
    \hline
	{} & \multicolumn{10}{|c|}{\textbf{Number of Keywords}} \\
	\textbf{System Name} & 10 & 20 & 30 & 40 & 50 & 60 & 70 & 80 & 90 & 100 \\
	\hline
	CLST-05 & 0.08 & 0.13 & 0.15 & 0.17 & 0.18 & 0.18 & 0.19 & 0.19 & 0.19 & 0.19 \\
	CLST-05B & 0.05 & 0.09 & 0.11 & 0.13 & 0.14 & 0.14 & 0.15 & 0.15 & 0.15 & 0.15 \\
	CLST-06 & 0.07 & 0.12 & 0.15 & 0.16 & 0.17 & 0.18 & 0.18 & 0.19 & 0.19 & 0.19 \\
	CLST-06B & 0.05 & 0.07 & 0.09 & 0.10 & 0.11 & 0.12 & 0.12 & 0.12 & 0.13 & 0.13 \\
	CLST-06NR & 0.08 & 0.12 & 0.14 & 0.16 & 0.17 & 0.18 & 0.18 & 0.19 & 0.19 & 0.19 \\
	CLST-06NRB & 0.05 & 0.08 & 0.09 & 0.11 & 0.11 & 0.12 & 0.12 & 0.13 & 0.13 & 0.13 \\
	\hline 
	MLT & & & & & & & 0.13 & & & \\
	\hline
  \end{tabular}
\end{table}

\setlength{\tabcolsep}{5pt}
\begin{table}
  \caption{Recall@100}
  \label{RECALL@100}
  \centering
  \begin{tabular}{|l|llllllllll|}
    \hline
	{} & \multicolumn{10}{|c|}{\textbf{Number of Keywords}} \\
	\textbf{System Name} & 10 & 20 & 30 & 40 & 50 & 60 & 70 & 80 & 90 & 100 \\
	\hline
	CLST-05 & 0.13 & 0.17 & 0.20 & 0.22 & 0.23 & 0.24 & 0.24 & 0.25 & 0.25 & 0.25 \\
	CLST-05B & 0.08 & 0.12 & 0.15 & 0.17 & 0.18 & 0.19 & 0.19 & 0.20 & 0.20 & 0.20 \\
	CLST-06 & 0.10 & 0.16 & 0.19 & 0.22 & 0.23 & 0.23 & 0.24 & 0.24 & 0.24 & 0.25 \\
	CLST-06B & 0.07 & 0.10 & 0.13 & 0.14 & 0.15 & 0.16 & 0.17 & 0.17 & 0.17 & 0.18 \\
	CLST-06NR & 0.11 & 0.16 & 0.19 & 0.21 & 0.22 & 0.23 & 0.24 & 0.25 & 0.25 & 0.25 \\
	CLST-06NRB & 0.07 & 0.11 & 0.13 & 0.15 & 0.16 & 0.16 & 0.17 & 0.17 & 0.18 & 0.18 \\
	\hline
	MLT & & & & & & & 0.17 & & & \\
	\hline
  \end{tabular}
\end{table}
The summary in Table~\ref{summary} clarifies which results are a statistically significant improvements over the MLT baseline (randomization test, *** means \(p<0.001\))
\setlength{\tabcolsep}{8pt}
\begin{table}
  \caption{Summary}
  \label{summary}
  \begin{tabular}{|l|ll|}
    \hline
	{} & \multicolumn{2}{|c|}{\textbf{Evaluation Metric}} \\
	\textbf{System Name} & RECALL@100 & PRES@100 \\
	\hline
	CLST-05 & \textbf{0.25} (***) & \textbf{0.19} (***) \\
	CLST-05 BOOST & 0.20 (***) & 0.15 (***) \\
	CLST-06	& 0.25 (***) & 0.19 (***) \\
	CLST-06 BOOST & 0.17 & 0.13 \\
	CLST-06 NO RETAG & 0.25 (***) & \textbf{0.19} (***)\\
	CLST-06 NO RETAG BOOST & 0.18 & 0.13 \\
	\hline
	MLT & 0.17 & 0.13 \\
	\hline
  \end{tabular}
\end{table}

The results show that BOOST is significantly decreasing performance. This can be interpreted as the scoring system having a good effect on selecting salient keywords over more general words, but not being able to catch the variations in relative importance of words in a way that is numerically in line with the boost factors of the search engine.

The correction of the POS tagging, which was tried only on the system CLST-06, does not generate a statistically significant improvement over the vanilla version. We analyze that the distortions on the parsing occur at different places than those where a specialization or combination occurs, which makes the system oblivious to this correction to a certain extent. Nonetheless, we’ll keep this correction in mind for future work, especially when additional features get extracted from the dependency parser.

The evaluation metrics keep increasing with the number of keywords, the difference being statistically non-significant between 80, 90 and 100 keywords. Our system overperforms the existing system, with a statistically significant improvement from 30 keywords onward. We keep the results based on 100 keywords as it is the setting where the best performance was delivered.

The significant result is that both CLST-05 and CLST-06 largely outperform the TFIDF-based baseline. Results show significant improvement in this setting of Query Generation, although the setting mixes the performance of the keyword extraction and the tweaking of the underlying search engine. Nonetheless, the approach of going away from term frequency methods to identify salient words in presence of an enforced writing style is proven to make sense. The term frequency allows for identification in absence of other information on how the text is written, while we can leverage the additional information that authors are restricted to deliver information in a way that is reflected in grammar, thus enabling us to work at the semantic level by working at the grammatical level.

\section{Conclusion}
In this work we used the sentence morphological features to identify keywords within patent claims, and used these keywords as query terms to retrieve other relevant patents. In this setting we establish a significative improvement over the existing baseline, based on term-frequency weighting methods.  \\
In the future we plan to apply and expand this work on other legal text sources (law, court decisions, contracts) with constrained writing style, should it be by regulation or tradition. We also see potential in adapting NLP tools that were designed or trained on conventional literature.

\section{Acknowledgments}
This work is a part of a paid internship offered by the European Patent Office. Opinions expressed in this paper are author’s only, and do not reflect the opinion of the European Patent Office.\\
The authors want to thank as well Volker H{\"a}hnke, Alexander Klenner-Bajaja, Stefan Klocke and Domenico Golzio, from the European Patent Office.

%
%
%
\bibliographystyle{splncs04}
\bibliography{epo.bib}

\end{document}